\documentclass{article}
\usepackage{graphicx,amssymb,amsthm}
\usepackage{url}
\usepackage[intlimits]{amsmath}
\usepackage{fullpage}

\newtheorem{theorem}{Theorem}

\newtheorem{lemma}[theorem]{Lemma}
\newtheorem{claim}{Claim}

\newtheorem*{lem3Rado}{Lemma 3 \cite{rado}}

\title{A Lower Bound on the Area of a 3-Coloured Disk Packing\footnote{A preliminary version of this paper was presented at the 19th Canadian Conference on Computational Geometry (CCCG~2007)~\cite{cccgpaper}.}}

\author{
        Peter Brass\thanks{Department of Computer Science,
        City College, CUNY, {\tt peter@cs.ccny.cuny.edu}}
		\and
		Ferran Hurtado\thanks{Departament de Matem\`atica Aplicada II, Universitat Polit\`ecnica de Catalunya
		{\tt ferran.hurtado@upc.edu}, partially supported by projects MEC MTM2006-01267 and Gen. Cat. 2005SGR00692}
		\and
		Benjamin Lafreniere\thanks{Department of Computer Science,
        University of Waterloo, {\tt bjlafren@cs.uwaterloo.ca}, partially supported by NSERC}
        \and
        Anna Lubiw\thanks{Department of Computer Science,
        University of Waterloo, {\tt alubiw@cs.uwaterloo.ca}, partially supported by NSERC}
		}

\index{Lafreniere, Benjamin}
\index{Lubiw, Anna}
\index{Brass, Peter}
\index{Hurtado, Ferran}

\begin{document}
\thispagestyle{empty}
\maketitle

\begin{abstract}
Given a set of unit-disks in the plane with union area $A$, what fraction of $A$ can be covered by selecting a pairwise disjoint subset of the disks?  Rado conjectured $1/4$ and proved $1/4.41$.  Motivated by the problem of channel-assignment for wireless access points, in which use of 3 channels is a standard practice, we consider a variant where the selected subset of disks must be 3-colourable with disks of the same colour pairwise-disjoint. For this variant of the problem, we conjecture that it is always possible to cover at least $1/1.41$ of the union area and prove $1/2.09$. We also provide an $O(n^2)$ algorithm to select a subset achieving a $1/2.77$ bound.
\end{abstract}

\section{Introduction}
\label{introduction}

\begin{figure}[h]
\begin{center}
\includegraphics[width=3in]{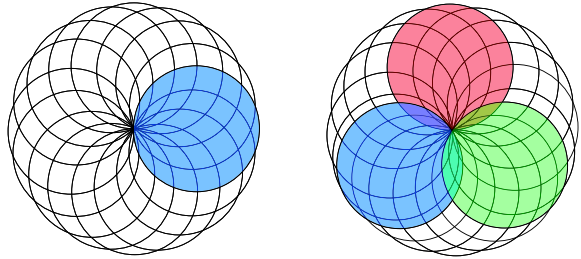}
\caption{Given a set of disks arranged in a circle with a very small mutual intersection, the most area we can cover is ${\frac{\pi}{4\pi}} = {\frac{1}{4}}$ with one colour (left) and approximately $\frac{1}{1.41}$ with 3-colours (right).}
\label{spirographs}
\end{center}
\end{figure}

Richard Rado studied the following problem: What is the largest $c_1$ such that, given any arrangement of unit-disks $D$ in the plane, we can always select a pairwise disjoint subset of disks that cover at least a fraction $c_1$ of the area of the union of $D$? Clearly $c_1 \leq {\frac{1}{4}}$, corresponding to the case shown in Figure~\ref{spirographs}~(left) where a large number of unit-disks share a very small intersection---the common intersection prevents us from selecting more than a single disk, while the union area of all disks approaches $4\pi$. Rado proved a lower bound of $c_1 \geqslant {\frac{\pi}{8\sqrt{3}}} \approx {\frac{1}{4.41}}$~\cite{rado} and conjectured the lower bound $c_1 \geqslant {\frac{1}{4}}$.

In this paper, motivated by the problem of channel-assignment for wireless access points, in which use of 3 channels is a standard practice, we focus on a variant of the above problem in which the disjointness constraint on the selected subset of disks is relaxed slightly. Given an arrangement of unit-disks $D$ in the plane, we want to find the largest $c_3$ such that we can always select and 3-colour a subset of the disks $C$ such that their union area covers at least a fraction $c_3$ of the union area of $D$, under the constraint that same-coloured disks must be pairwise disjoint. We will prove that, for any given arrangement of unit-disks, $c_3 \gtrapprox {\frac{1}{2.09}}$. This result is stated formally in Theorem~\ref{weightedthm}.

\newcommand{\weightedbound}{
Let $D$ be a collection of unit-disks in the plane with union area $A$. For $C$ a 3-coloured subset of $D$ with same-coloured disks pairwise disjoint, let $A_C$ denote $C$'s union area. We can always select a $C$ such that ${\frac{A_C}{A}} \gtrapprox {\frac{1}{2.09}}$ and thus $c_3 \gtrapprox {\frac{1}{2.09}}$.
}
\begin{theorem} \label{weightedthm}
\weightedbound
\end{theorem}

Note that $c_3 \leq {\frac{1}{1.41}}$, corresponding to selecting three minimally pairwise intersecting disks in an arrangement as shown in Figure~\ref{spirographs}~(right). We conjecture that this bound is achievable for any arrangement of unit-disks.

Before presenting our proof of Theorem~\ref{weightedthm} we will review Rado's proof and prove the weaker result that $c_3 \gtrapprox {\frac{1}{2.77}}$, as formally stated in Theorem~\ref{basicthm}. We include this proof as a stepping stone to Theorem~\ref{weightedthm} and because it leads to an $O(n^2)$ algorithm for the disk selection problem.

\newcommand{\basicbound}{
Let $D$ be a collection of unit-disks in the plane with union area $A$. For $C$ a 3-coloured subset of $D$ with same-coloured disks pairwise disjoint, let $A_C$ denote $C$'s union area. We can always select a $C$ such that ${\frac{A_C}{A}} \gtrapprox {\frac{1}{2.77}}$.
}
\begin{theorem} \label{basicthm}
\basicbound
\end{theorem}

The rest of this paper is organized as follows. In Section~\ref{motivation} we present our motivation for exploring this problem and discuss some related problems. In Section~\ref{thebasicbound} we review Rado's proof and prove Theorem~\ref{basicthm}. In Section~\ref{weightedsection} we prove Theorem~\ref{weightedthm}. In Section~\ref{algorithmstuff} we present an $O(n^2)$ time algorithm to select a subset which achieves our $1/2.77$ bound 
from Theorem~\ref{basicthm} and prove that 
the lattice-positioning step in our algorithm 
is 3SUM-hard. 
Finally, Section~\ref{extension} discusses bounds for $k$-colours when $k \ne 3$.

\section{Motivation and Related Work}
\label{motivation}

In this section we describe a wireless network deployment problem that motivates our 3-coloured disk packing problem, and discuss other disk packing results related to ours.

Wi-Fi (IEEE 802.11) wireless networks are becoming a ubiquitous feature in modern businesses, universities, parks, etc. In a typical Wi-Fi deployment scenario, a set of candidate locations are determined for wireless access points (APs). A subset of the candidate locations must be chosen along with a channel assignment for each installed AP in order to maximize the area covered by the wireless network while minimizing interference. Interference occurs when two  APs using the same channel are within range of one another, preventing users in range of both APs from communicating with either. For a more in depth discussion of wireless network deployment, see the introduction to~\cite{keshav}. We make the simplifying assumption that the coverage area of each AP is a disk, though in practice coverage areas may be more irregular. We also assume that every AP uses the same power level---i.e.~that all the disks have the same radius. This is not entirely unjustified, as the problem of power control for Wi-Fi network devices is a complex research area in its own right~\cite{powercontrol}. Finally, the set of potential locations for APs is limited to those that are physically possible and aesthetically acceptable. We model this as a finite set of disk center points. Under these conditions, the wireless network deployment problem becomes a $k$-coloured disk packing problem, where $k$ is the number of allowable channels. To justify $k = 3$, we will discuss the constraints on channel selection in Wi-Fi networks.

The IEEE 802.11 standard establishes a number of requirements on the radio frequency characteristics of 802.11 hardware. The 2.4-Ghz band used by 802.11 devices is broken into 11 channels for the North American domain and 13 channels for the European domain. Unfortunately, the number of effective channels is much lower due to interference between channels with center frequencies close to one another. A study by wireless hardware maker Cisco Systems recommends that wireless network deployments only use three channels (1, 6, and 11 for the North American domain), finding that even a four-channel scheme can cause unacceptable degradation of service in systems with a high volume of users~\cite{cisco}.

A more sophisticated formalization of the deployment problem allows disks assigned to the same channel to overlap but only counts the area where there is no interference---i.e. the area of the set of points covered by only one disk on some channel. In terms of colouring, the problem is to colour a subset of the given disks to maximize the area of $\{ p \in \mathbb{R}^2 \mid $ for some colour, point $p$ is in exactly one disk of that colour$\}$. We call this the \emph{1-covered area}. Asano et al.~\cite{asanobrasshalftoning} proved that it is always possible to achieve approximately $\frac{A}{4.37}$ 1-covered area using only one colour. This model has also been considered with respect to two other optimization problems. For the problem of maximizing the 1-covered area using one colour, previous work has focused on approximation algorithms (though no proof yet exists, it is suspected that this problem is NP-hard). In \cite{asanobrasshalftoning}, Asano et al. present a 5.83-approximation algorithm with polynomial runtime. In \cite{ptasdisccovering}, Chen et al. show that the problem admits a polynomial time approximation scheme when the radius of the largest disk over the radius of the smallest disk is a constant. 

Another well-explored optimization problem is \emph{conflict-free colouring}---here the goal is to minimize the number of colours needed to 1-cover the whole area, i.e.~the union of the given disks. Even et al.~\cite{evenlogn} prove that $O(\log n)$ colours are always sufficient and sometimes necessary for any given disks of general radii. Alon et al.~\cite{conflictfreeshallow} have shown that, if each disk intersects at most $k$ others, then $O(\log^3 k)$ colours are sufficient for a conflict-free colouring, improving the bound from \cite{evenlogn} when $k$ is much smaller than $n$. There is also work on online algorithms for conflict-free colouring~\cite{fiatonlinecf}, and on conflict-free colouring of regions other than disks~\cite{cfsimpleregions}.

\section{The Basic Bound}
\label{thebasicbound}
\subsection{Rado's Proof}
\label{lattices}

\begin{figure}[h]
\begin{center}
\includegraphics[width= 1.2in]{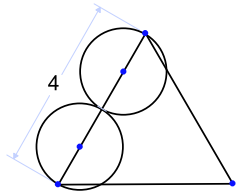}
\caption{A triangular lattice ensures that selected disks are pairwise disjoint.}
\label{twocirclesintriangle}
\end{center}
\end{figure}

The idea behind our proof of the basic bound is similar to that used by Rado in~\cite{rado}. In Rado's proof, a regular triangular lattice of side length 4 is positioned over the set of disks $D$ and for each lattice point that falls in $\cup D$ one disk containing that lattice point is selected. The side length of the lattice guarantees that disks selected in this manner will be pairwise disjoint (see Figure~\ref{twocirclesintriangle}). Thus, supposing we can prove a lower bound of $k$ on the number of lattice points intersecting any given set of disks, we immediately obtain a lower bound of $c_1 \geqslant \frac{\pi k}{A}$. We get such a lower bound by applying Lemma~\ref{radofundamentalcell} which states that we can position the lattice to contain at least ${\frac{A}{2\alpha}}$ points in $\cup D$, where $\alpha$ is the area of a triangle in the lattice. Since $\alpha = 4\sqrt{3}$ for our lattice, we can position the lattice to contain at least $k = \frac{A}{8\sqrt{3}}$ points in $\cup D$ and therefore $c_1 \geqslant {\frac{\pi}{8\sqrt{3}}}$.

\begin{lem3Rado}
\refstepcounter{theorem} 
\label{radofundamentalcell}
Given a region of the plane $G$ with area $A$, and a triangular lattice in which each triangle has area $\alpha$, the lattice can always be positioned such that it contains ${\frac{A}{2\alpha}}$ points in $G$.
\end{lem3Rado}
\begin{proof}
Rado's proof~\cite{rado} uses the concept of the ``fundamental cell'' of a lattice---for a regular triangular lattice the fundamental cell $F$ consists of a pair of adjacent triangles (see Figure~\ref{3coloursgrid}). Given an arbitrary placement of the lattice, each triangle of the lattice can be translated to $F$ along with whatever part of $G$ is contained in the triangle. The translated parts of $G$ may ``overlap'' in $F$---there may be points on the fundamental cell intersecting multiple translated portions of $G$. Supposing a point $p$ in $F$ intersects $k$ translated portions of $G$, then repositioning the lattice such that $p$ is a lattice point ensures that $k$ lattice points intersect $G$. In this case, we refer to $k$ as the \emph{depth} of point $p$ in the fundamental cell. Since the area of $G$ is $A$, the total area of all portions of $G$ translated to $F$ is clearly also $A$. Therefore, we have portions of $G$ with total area $A$ translated to a region of area $2\alpha$ and a point of depth at least ${\frac{A}{2\alpha}}$ must exist.
\end{proof}

\subsection{A Bound of $c_3 \gtrapprox 1/2.77$}
\label{proofoftheorem}

\setcounter{theorem}{1} 
\begin{theorem}
\basicbound
\end{theorem}
\setcounter{theorem}{3} 

\begin{figure}[h]
\begin{center}
\includegraphics[width=1.2in]{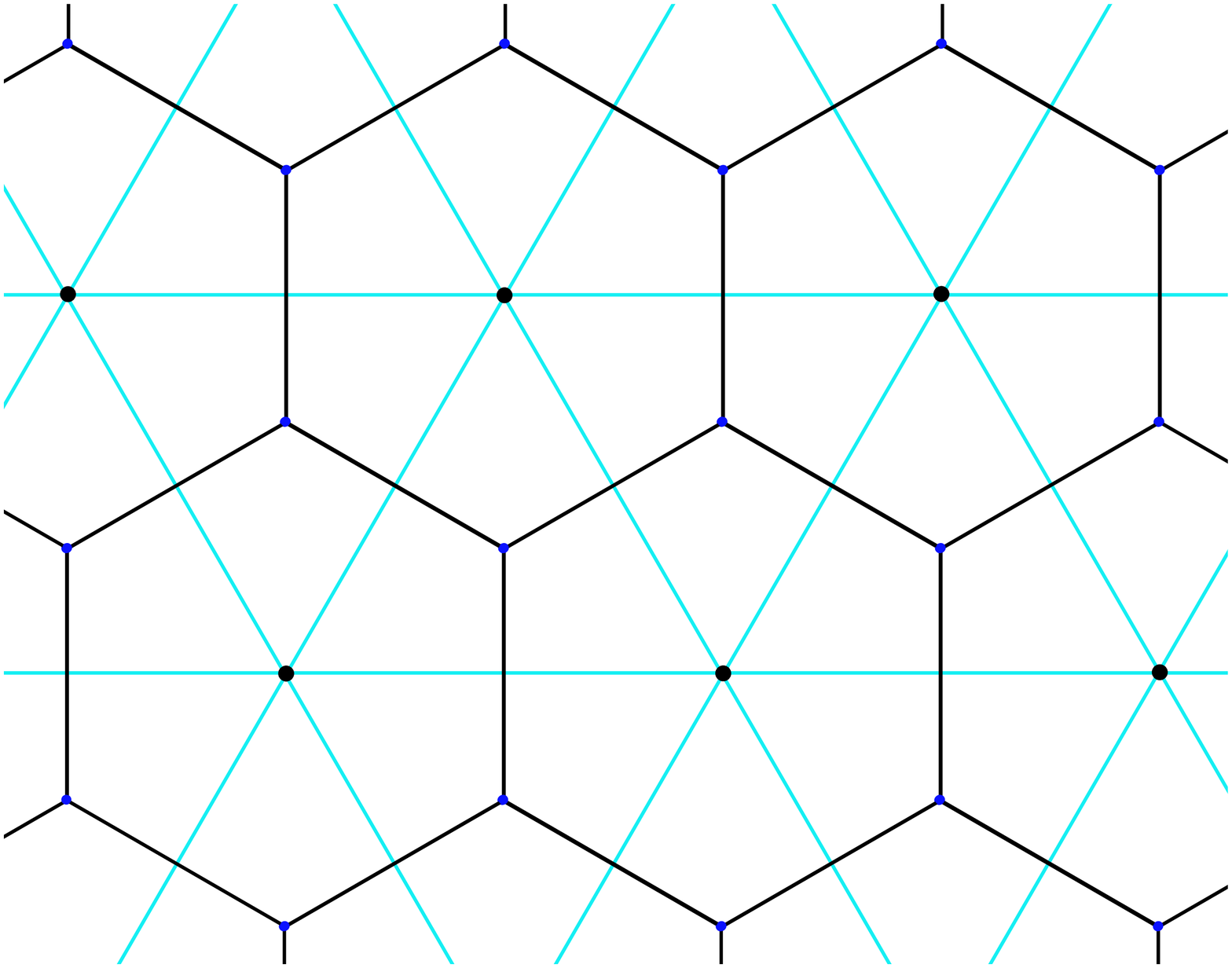}
\caption{The Voronoi tessellation of the triangular lattice points forms a grid of regular hexagonal cells.}
\label{tessellation}
\end{center}
\end{figure}

\begin{proof}
To solve our variation of the problem we use a finer triangular lattice with side length $\frac{4\sqrt{3}}{3}$. The points of the lattice are 3-coloured such that no two lattice points of the same colour are adjacent. For any placement of the lattice, select a subset $C$ of $D$ as follows: for each lattice point $p$ in the union of $D$, select a disk containing $p$ and assign the disk the colour of $p$. The side length of the lattice ensures that no disk contains two lattice points so the selection and colouring are well-defined. It also ensures that disks assigned the same colour are pairwise disjoint (see Figure~\ref{3coloursgrid}).

\begin{figure}[h]
\begin{center}
\includegraphics[width= 2in]{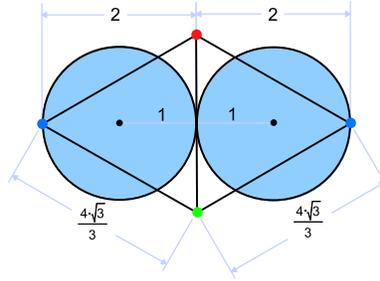}
\caption{A finer 3-coloured lattice ensures only that same-coloured selected disks are pairwise disjoint.}
\label{3coloursgrid}
\end{center}
\end{figure}

Also observe that, by Lemma~\ref{radofundamentalcell}, we can position the lattice to intersect the union area of $D$ in at least $\frac{A\sqrt{3}}{8}$ points, so $|C| \geqslant {\frac{A\sqrt{3}}{8}}$. While same-coloured disks in $C$ are pairwise disjoint, differently coloured disks may not be, so $|C| \pi$ is only an upper bound on $A_C$.

To derive a lower bound we will partition the union of $C$ using the triangular lattice's Voronoi tessellation which has regular hexagonal cells of side length $\frac{4}{3}$ and vertices at the barycenters of the triangular lattice (see Figure \ref{tessellation}). Suppose disk $d \in C$ contains lattice point $p$ which lies in hexagonal cell $h$. If we count only the area of $d \cap h$, and sum over all $d$, this gives a lower bound on $A_C$. Thus if we establish a lower bound $\Delta$ on the minimum possible area of $d \cap h$ then $A_C \geqslant |C| \Delta \geqslant \frac{A\sqrt{3}}{8}\Delta$. In Lemma~\ref{lemma1}, which we will prove in Section~\ref{mindiskhexint}, we show that $\Delta \approx 1.6645$. From the lower bound on $A_C$ we reach our desired lower bound on $c_3$ of
\small
\begin{align*}
c_3 \geqslant {\frac{A_C}{A}} = {\frac{\sqrt{3}}{8}}\Delta \approx \frac{1}{2.77}
\end{align*}\end{proof}
\normalsize

\subsection{Minimum Disk-Hexagon Intersection}
\label{mindiskhexint}

\begin{lemma}
\label{lemma1}
Given a regular hexagon $h$ with center point $X$ and side length $\frac{4}{3}$, and any unit-disk $d$ containing point $X$, the minimum
area of intersection $\Delta$ between $h$ and $d$ is approximately $1.6645$, or more precisely
\small
\begin{eqnarray*}
\Delta &=& {\frac{\sqrt{3}}{36}}\,+{\frac{\sqrt{11}}{12}}\,+{\frac{\pi}{2}}\, -{\frac{1}{2}}\,\arctan \left( {\frac{5\,\sqrt {3}-\sqrt {11}}{5+\sqrt {11}\sqrt {3}}} \right)\\
\end{eqnarray*}
\end{lemma}
\normalsize

\begin{proof}
We use elementary geometry to argue that the minimum area of intersection is achieved by a disk $d$ with $X$ on its boundary. Then, parameterizing by the angle $\theta$ between the horizontal axis and the ray from $X$ to the center of $d$, we use the symbolic geometry package Geometry Expressions to give a formula for the area of intersection and use Maple to compute 0's of the first derivative, finding that the minimum is as stated above, and occurs in the configuration shown in Figure~\ref{tightboundtri}. Further details are included in Appendix~\ref{theappendix}.
\end{proof}

Our proof of Lemma~\ref{lemma1} also shows that the lower bound $A_C \ge |C| \Delta$ is tight, as shown by the example in Figure~\ref{tightboundtri} where the union of $C$ is exactly partitioned by the hexagons and each disk intersects its hexagon in the minimum area $\Delta$. However, note that this does not mean that our bound on $c_3$ is tight. In particular, for the example shown in Figure~\ref{tightboundtri} we can capture the whole area by 3-colouring the disks.

\begin{figure}[h]
\begin{center}
\includegraphics[width=1.2in]{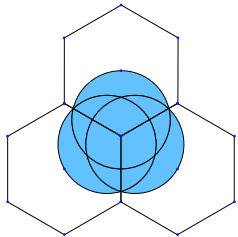}
\caption{In this arrangement of selected disks, the lower bound on the contribution of each disk is realized.}
\label{tightboundtri}
\end{center}
\end{figure}

\section{Deriving a Better Bound}
\label{weightedsection}

In our proof of Theorem~\ref{basicthm} we only counted the minimum intersection of a disk with its selecting Voronoi cell (the Voronoi cell of the lattice point that selects the disk). Suppose we commit to this accounting scheme. We can clearly maximize the intersection of the disk with its Voronoi cell by centering the selecting lattice point in the disk. This suggests that we can use a weighting function that prefers placing a lattice point in the center of a disk, 
and do a more intricate analysis of the contribution of each disk.

\begin{figure}[h]
\begin{center}
\includegraphics[width=2.2in]{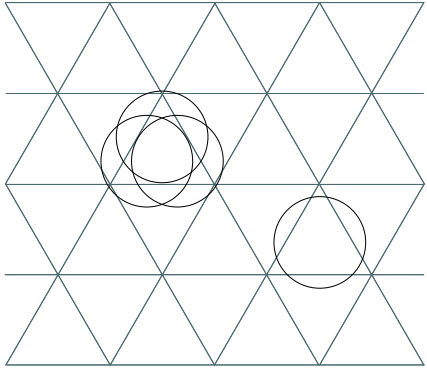}
\caption{The lattice positioning shown selects the maximum number of disks, however an alternate lattice positioning can select fewer disks and cover more area.}
\label{moreisnotbetter}
\end{center}
\end{figure}

Looking at this from another direction, in the proof of Theorem~\ref{basicthm} we optimized the number of disks selected, rather than the area of the intersection between selected disks and their selecting Voronoi cells. This approach can be improved because, among all subsets of disks that can be selected using a lattice, the largest subset of disks does not necessarily cover the largest area (e.g.~see Figure~\ref{moreisnotbetter} where selecting the three intersecting disks using the lattice shown gives a subset with union area $3\Delta \approx 4.99$, while an alternate lattice positioning that selects two disjoint disks gives a subset with union area $2\pi \approx 6.28$).

The above points suggest that we can improve our bound on $c_3$ by using a more sophisticated criterion for lattice positioning based on the area contributed by selected disks rather than the number of disks selected. We will use this approach to prove Theorem~\ref{weightedthm}.

\subsection{A Bound of $c_3 \gtrapprox 1/2.09$}
\label{proofofweightedbound}

\setcounter{theorem}{0} 
\begin{theorem}
\weightedbound
\end{theorem}
\setcounter{theorem}{4} 

\begin{proof}
To prove this better bound, we use a weighted version of Lemma~\ref{radofundamentalcell}. Whereas Lemma~\ref{radofundamentalcell} proved that a lattice $L$ can be positioned to intersect $\cup D$ in at least a certain number of points, we want to prove that $L$ can be positioned to intersect $\cup D$ with at least a certain weight. 

We now define the weight function. For point $p$, let $H(p)$ be the regular hexagon of side length $\frac{4}{3}$ centered at $p$. Thus $H(p)$ is the Voronoi cell of $p$ if our triangular lattice is translated to include point $p$. For $p \in \cup D$, let $d(p)$ be the disk containing $p$ whose intersection with $H(p)$ has maximum area. Now let $w(p) = area(H(p) \cap d(p))$ for $p \in \cup D$, $w(p) = 0$ otherwise. Then $w(p)$ measures the area contributed by including $p$ in the lattice (given our method of choosing disks based on the lattice, and our accounting scheme of counting only the area of the disk in the Voronoi cell).

We want to choose a lattice $L$ to maximize $W(L) = \sum_{p \in L} w(p)$. 
Proving a lower bound on the maximum value of $W(L)$ for lattices of the type used in our proof of Theorem~\ref{basicthm} will give us a lower bound on $c_3$.

As in the proof of Lemma~\ref{radofundamentalcell} we consider translating portions of $\cup D$ onto a fundamental cell. After translation, each point $p$ on the fundamental cell is assigned the sum of the weights of $p$'s intersection with each translated portion of $\cup D$. Supposing that a point $p$ on the fundamental cell has weight $m$, then positioning $L$ such that $p$ is a lattice point ensures that $W(L) = m$. Thus, if we can prove that the total weight of the portions of $\cup D$ translated to the fundamental cell is $B$, then a point in the fundamental cell (and therefore a lattice positioning) with weight at least $\frac{\sqrt{3}}{8}B$ must exist.

\begin{figure}[h]
\begin{center}
\includegraphics[width=1.2in]{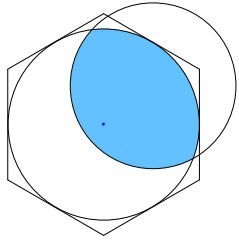}
\caption{Considering the intersection between a selected disk and the largest disk that can be contained in the disk's Voronoi cell provides a simple lower bound for $w(p)$ in terms of distance from $p$ to the nearest disk center.}
\label{simplerintfunction}
\end{center}
\end{figure}


We cannot compute exactly the value of $B$, which is the integral of $w(p)$ over $p \in \cup D$, but we can obtain a lower bound. Our first step is to replace the true weight function $w(p)$ by a lower bound, $w_l(p)$, in which we replace the hexagon by its inscribed circle (see Figure~\ref{simplerintfunction}). Specifically, $w_l(p)$ is the area of the intersection of two discs: the disk of radius $2 / \sqrt{3} $ centered at $p$, and the unit-disk in $D$ whose center is closest to $p$. Note that $w_l(p)$ depends only on the distance, $r$, from $p$ to the nearest disk center in $D$. We will overload the notation and define $w_l(r)$, for $r \in [0,1]$, to be the area of the intersection of a unit-disk and a disk of radius $2 / \sqrt{3} $ whose centers are distance $r$ apart. We note that $w_l(r)$ is a continuous, decreasing function, and that we can write an explicit formula for it:

\small
\begin{align*}
w_l(r) = \begin{cases}
		\pi & \text{if $0 \leq r \leq \frac{2}{\sqrt{3}} - 1$}\\\\
		\arccos \left(\frac{1}{2}\,{\frac {{r}^{2}-\frac{1}{3}}{r}} \right) +\frac{4}{3}\,\arccos\left( \frac{1}{4}\,{\frac { \left( {r}^{2}+\frac{1}{3} \right) \sqrt {3}}{r}}\right)-\\ \frac{1}{2}\,\sqrt { \left( -r+1+\frac{2}{\sqrt{3}}\right)  \left( r+1-\frac{2}{\sqrt{3}}\right)  \left( r-1+\frac{2}{\sqrt{3}}\right)  \left( r+1+\frac{2}{\sqrt{3}}\right) } & \text{if $\frac{2}{\sqrt{3}} - 1 < r \leq 1$}\\
	\end{cases}
\end{align*}
\normalsize

We will express $B$ as an integral in terms of variable $r$. Note that the points on the boundary of $\cup D$ are precisely the points at distance 1 from the closest disk center. More generally, we can capture the points that are distance $r$ from the closest disk center as follows. For unit-disk $d \in D$, let $d_r$ be a disk of radius $r$ at the same center. Let $D_r$ be $\{d_r : d \in D\}$. Then the points that are distance $r$ from the closest disk center are precisely the points on the boundary of $\cup D_r$. Let $p(r)$ be the length of the boundary of $\cup D_r$. As $w(r) \geq w_l(r) \geq 0$, we obtain:

\begin{align*}
B &= \int_0^1 p(r)w(r)\,dr \geq \int_0^1 p(r)w_l(r)\,dr
\end{align*}

We prove a lower bound on the latter integral that eliminates $p(r)$.

\begin{lemma}
\label{integral-bound}
$B = \int_0^1 p(r)w(r)\,dr \geq \int_0^1p(r)w_l(r)\,dr \geq 2A\int_0^1rw_l(r)\,dr$.
\end{lemma}

After this we are done: we plug in the expression for $w_l(r)$ and evaluate the integral using Maple to obtain a lower bound on $B$ of $2.207 A$. Therefore a lattice positioning with weight at least $\frac{\sqrt{3}}{8}B \gtrapprox \frac{A}{2.09}$ must exist, and we get our desired lower bound of $c_3 \gtrapprox \frac{1}{2.09}$.

It remains to prove the Lemma.

\begin{proof}[Proof of Lemma~\ref{integral-bound}]

We want to prove $\int_0^1p(r)w_l(r)\,dr \geq 2A\int_0^1rw_l(r)\,dr$, equivalently:

\begin{align*}
\int_0^1[p(r) -2rA] w_l(r)\,dr \ge 0
\end{align*}

Let $A(r)$ be the area of $\cup D_r$, which is related to the perimeter by the fact that $p(r) = A'(r)$. Define $f(r) = A(r) - r^2 A$.  Recall that $A = A(1)$ is the area of $\cup D$. Note that $f(0) = f(1) = 0$. Now $f'(r) = A'(r) - 2rA = p(r) - 2rA$, and the inequality we want to prove is transformed into $\int_0^1 f'(r) w_l(r)\,dr \ge 0$.

We apply integration by parts, noting that $f'$ is continuous, and that $w_l'$ is continuous except at $\frac{2}{\sqrt{3}} - 1$.

\begin{align*}
\int_0^1f'(r)w_l(r)\,dr \ \ &= \ \  
f(r)w_l(r)\biggl|_0^1  \ - \ \int_0^1 f(r) w_l'(r)\,dr
\ \ = \ \ - \int_0^1 f(r) w_l'(r)\,dr
\end{align*}

We will prove below that $A(r) \ge r^2 A$.  Thus $f(r) \ge 0$ for all $r \in [0,1]$.  Now $w_l(r)$ is a decreasing function, so $w_l'(r) \le 0$ for all $r \in [0,1]$.  
The integral of a negative function is negative, and this completes the proof.

\end{proof}

\end{proof}

\begin{figure}[h]
\begin{center}
\includegraphics[width=3in]{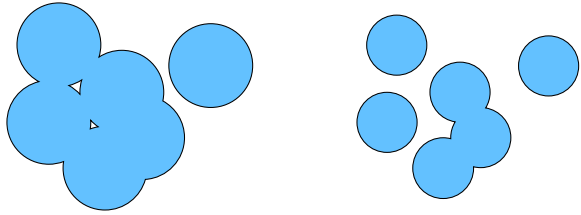}
\caption{The union of a set of disks before (left) and after a radius scaling has been applied (right).}
\label{scalingvisual}
\end{center}
\end{figure}

\begin{claim}
\label{bollobasextended}
Given a collection of unit-disks with union area $A$, if we scale the radius of each disk by $r \in [0,1]$ then the union area of the scaled disks will be at least $r^2A$. 
\end{claim}
\begin{proof}
We want to scale the radius of each disk by $r$. We accomplish this in two steps. First we scale the whole plane by a factor of $r$. This reduces the area to $r^2A$. It also decreases the distance between the centers of any two disks by a factor of $r$.  The second step is to translate each scaled disk back to its original position. During the translation the distance between the centers of any two disks increases continuously. By a result of Bollob\'{a}s~\cite{bollobas}, the union area of a set of congruent disks moving apart from one another continuously cannot decrease, and therefore the area of the final arrangement of scaled disks is at least $r^2A$.
\end{proof}

\section{Algorithm}
\label{algorithmstuff}

In this section we give an $O(n^2)$ time algorithm to select and 3-colour a subset $C$ of a set $D$ of unit-disks so that the area bound given in Theorem~\ref{basicthm} is realized. The proof of our theorem is constructive, and the only algorithmic issue is positioning the lattice so that at least $\frac{A\sqrt{3}}{8}$ lattice points are in $\cup D$ (the union of all disks in $D$). We give an $O(n^2)$ time algorithm for this lattice positioning problem. We also prove that the lattice positioning problem is 3SUM-hard, providing evidence that an $O(n^2)$ time algorithm is the best we can expect with this approach.

To solve the lattice positioning problem we follow the constructive approach used to prove Lemma~\ref{radofundamentalcell}, initially positioning the lattice arbitrarily and then translating all triangles in the lattice along with portions of $\cup D$ to a fundamental cell $F$. Thus the one remaining detail is how to capture the translated portions of $\cup D$ so that we can compute a point of maximum depth. Our basic idea involves translating all of the disks and then computing and traversing their arrangement. Each disk $d$ intersects at most 4 translates of $F$. We make 4 translated copies of $d$, and record which translate of $F$ they come from. Computing this set of translated disks $D'$  takes $O(n)$ time.  Computing the arrangement of $D'$,  ${\cal A}(D')$, takes $O(n^2)$ time using the incremental insertion algorithm of Chazelle and Lee~\cite{chazelleleecircleplacement}.

It is easy to traverse ${\cal A}(D')$ to compute maximum depth in $D'$---the depth increases when we enter a disk and decreases when we exit.   However, this is not quite what we want; we want depth with respect to $\cup D$ translated to $F$, which is different from depth in $D'$ due to disks that overlap originally in $D$. Our solution is to traverse ${\cal A}(D')$ maintaining the depth $c_i$ in each translate $i$ of $F$.  Note that there are $O(n)$ translates of $F$ that intersect $\cup D$. We also maintain a count $c$ of the number of non-zero $c_i$'s. The maximum value of $c$ over cells of ${\cal A}(D')$ gives us what we want.

We now prove that the lattice positioning problem discussed above is 3SUM-hard. A problem is \emph{3SUM-hard} if it is harder than the problem of determining whether a set $S$ of $n$ integers contains three elements $a, b, c \in S$ such that $a + b + c = 0$. The best known algorithms for this problem take $O(n^2)$ and it is an open problem to do better \cite{go95}.

\begin{theorem}
\label{3sumproof}
The following problem is 3SUM-hard: Given an integer $k$, real number $s$, and a set $D$ of $n$ unit-disks in the plane, determine whether a triangular lattice of side length $s$ can be positioned such that it intersects the union area of $D$ in at least $k$ points.
\end{theorem}

\begin{proof}
We show that our problem is harder than the known 3SUM-hard problem of determining whether there is a point of depth $k$ in a set of unit radius disks in the plane. The more general problem for variable radius disks is proved 3SUM-hard in \cite{diskdepth3sum} and the reduction is easily modified to produce unit radius disks.

Our reduction is as follows.  Given a set $D$ of unit radius disks in the plane, place an equilateral triangle $T$ large enough to contain all of $D$. Expand $T$ to a triangular lattice, and translate each disk of $D$ to a different cell in the lattice, with the same orientation. Let the translated set of disks be $D'$. Then there is a point of depth $k$ in $D$ if and only if the lattice can be translated to intersect $D'$ in $k$ points. This reduction takes linear time.
\end{proof}

\section{Preliminary Bounds for $k$-Colours, $k \ne 3$, and Future Work}
\label{extension}

While we have focused on the case of 3 colours, since that is relevant for channel assignment in wireless networks, it is interesting to see what bounds can be derived for other numbers of colours. For 2 colours, we can use the techniques used to prove Theorems~1~and~2 but with a 2-coloured square lattice of side length $2\sqrt{2}$. This approach leads to a bound of $c_2 \gtrapprox \frac{1}{2.82}$ analogous to the bound from Theorem~\ref{weightedthm}, and a weaker bound of $c_2 \gtrapprox \frac{1}{3.37}$ for which an $O(n^2)$ algorithm exists for selecting disks to satisfy the bound. Full details are included in the thesis of Benjamin Lafreniere~\cite{bensthesis}.

For the more general problem of $k$-colours, Theorem~\ref{kcolourbound} presents some preliminary results, demonstrating a bound for all $k$ such that  $k \in \{ i^2 + ij + j^2 \mid i, j \in \mathbb{N}\}$. The number of such $k$ lower than a given $x \in \mathbb{N}$ is given by $\Theta(\frac{x}{\sqrt{\log x}})$, so the set of such $k$ is thin (density 0).

\begin{theorem}
\label{kcolourbound}
Given $k$ colours, where $k \in \{ i^2 + ij + j^2 \mid i, j \in \mathbb{N}\}$ we can select and colour a subset of disks such that same-coloured disks are disjoint and their union area covers at least $A\frac{1}{(1 + \delta_{k})^2}$ where $\delta_{k} = \frac{2}{\sqrt{3}}\left( \frac{2}{\sqrt{k} - \frac{2}{\sqrt{3}}} \right)$.
\end{theorem}

\begin{proof}
For all $k \in \{ i^2 + ij + j^2 \mid i, j \in \mathbb{N}\}$, distance $\sqrt{k}$ occurs in the unit triangular lattice $L$, and by $\frac{\pi}{3}$ rotational symmetry, an entire sublattice with side length $\sqrt{k}$ exists. Thus we can partition $L$ into $k$ triangular lattices of side length $\sqrt{k}$ and assign each a unique colour. We then scale $L$ such that distance 2 separates the enclosing disks of Voronoi cells of same-coloured lattice points by applying a scaling factor of $\alpha_{k} = \frac{2}{\sqrt{k} - \frac{2}{\sqrt{3}}}$. Now, each disk in $D$ is assigned to the Voronoi cell containing its center. We select from each Voronoi cell one associated disk (if there are any) and colour it to match the Voronoi cell's lattice point. Note that by our scaling, same-coloured selected disks cannot intersect.

If a point $p$ is in $\cup D$ but is not in any selected disk, then the disk covering $p$ intersects another disk with center in the same Voronoi cell, and the distance between their center points is less than the diameter of the Voronoi cell $\delta_{k} = \frac{2}{\sqrt{3}}\alpha_{k}$. Now, if all selected disks were blown up by a factor of $1 + \delta_{k}$, $p$ would be covered by some selected disk and the union of selected disks would cover at most $A(1 + \delta_{k})^2$. Thus, if we allow $k$ colours, we can cover at least $A\frac{1}{(1 + \delta_{k})^2}$.
\end{proof}

\section*{Acknowledgements}
\label{acknowledgements}

This problem was introduced to us by S.~Keshav, University of Waterloo, and the work was initiated at the 5th McGill-INRIA Workshop on Computational Geometry in Computer Graphics at McGill's Bellairs Research Institute in 2006. The workshop was organized by Hazel Everett, Sylvain Lazard, and Sue Whitesides. We thank S.~Keshav, and the participants of the McGill-INRIA workshop for fruitful discussions. We thank Ross Willard, University of Waterloo, for advice on several issues in Section~\ref{weightedsection}.



\bibliographystyle{abbrv}

\newpage
\normalsize
\appendix
\section{Proof of Lemma~\ref{lemma1}}
\label{theappendix}

\begin{figure}[h]
\begin{center}
\includegraphics[width=2.5in]{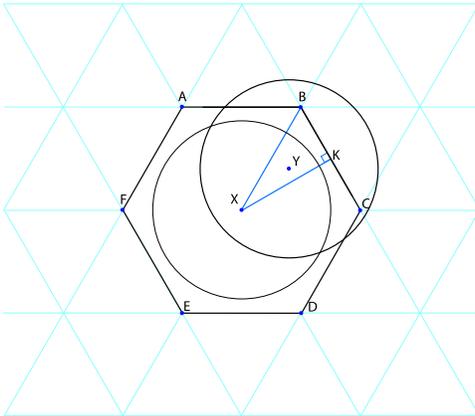}
\caption{Sliding disk $d$ outward from the center of $h$ results in a smaller intersection between $d$ and $h$}
\label{slide_outward_proof}
\end{center}
\end{figure}

Our first claim is that the minimum area of intersection is achieved by a disk $d$ positioned such that $X$, the center point of hexagon $h$, lies on its boundary. Suppose this is not the case. By symmetry, it suffices to consider possible placements of $Y$, the center point of $d$, within the intersection of wedge $BXK$ and the unit-disk centered at $X$ in Figure~\ref{slide_outward_proof}. For any position of $Y$, moving $Y$ to the right along a line parallel to $AB$ decreases the area of intersection, since the portion of $d - h$ lying above the supporting line of $AB$ stays the same, and the portion of $d - h$ below the supporting line of $AB$ strictly increases (by containment). Thus we can move $Y$ to the right until it lies either on $XK$ or the boundary of the disk centered at $X$. For $Y$ on $XK$, moving $Y$ toward $K$ decreases $d \cap h$ because when the diameter of $d$ parallel to $BC$ lies strictly inside $h$, the area of $d - h$ increases (by containment), and when the diameter is not strictly contained in $h$, the area of $d \cap h$ decreases (by containment).

\begin{figure}[h]
\begin{center}
\includegraphics[width=2in]{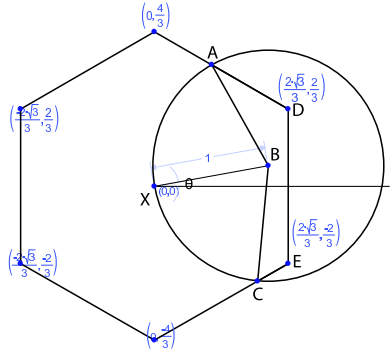}
\caption{Calculating the intersection area as a function of angle $\theta$ (Case 1).}
\label{intersection_on_edge}
\end{center}
\end{figure}

\begin{figure}[h]
\begin{center}
\includegraphics[width=1.9in]{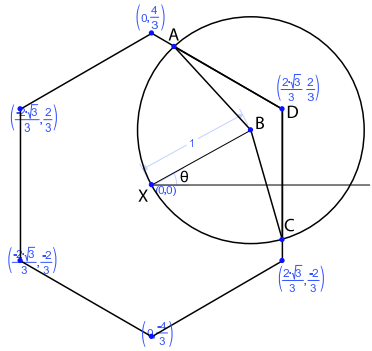}
\caption{Calculating the intersection area as a function of angle $\theta$ (Case 2).}
\label{intersection_on_point}
\end{center}
\end{figure}

Thus we can restrict our attention to the minimum area of intersection with $h$ among disks whose boundary contains point $X$. To find this minimum, we assume that $X = (0,0)$ and express the area of intersection $f(\theta)$ in terms of angle $\theta$ between the center of a disk $d$, the center $X$ of $h$, and the $x$-axis. There are two general cases to consider. Case~1 occurs when $d$ contains two vertices of $h$ (e.g.~Figure~\ref{intersection_on_edge}). Case~2 occurs when $d$ only contains a single vertex of $h$ (e.g.~Figure~\ref{intersection_on_point}). Note that in either case we can express the area of intersection as the sum of the area of a polygon and a circle sector. For instance, in Figure~\ref{intersection_on_edge} the area of intersection is the sum of the area of polygon $ABCED$ and the area of the sector of $d$ interior to angle $ABC$.

By symmetry, we need only consider the area of intersection for $0 \leq \theta \leq \frac{\pi}{6}$. As a result, we can use the cases shown in Figures~\ref{intersection_on_edge} and \ref{intersection_on_point} to derive a formula for $f(\theta)$. Specifically, we use the symbolic geometry package Geometry Expressions to derive formulas relating $\theta$ and the intersection points between the boundaries of $d$ and $h$ (i.e. points $A$ and $C$ in Figures~\ref{intersection_on_edge} and \ref{intersection_on_point}). The derived formulas for these points, along with formulas for the other points in Figures~\ref{intersection_on_edge} and \ref{intersection_on_point} are given in Appendix~\ref{appendix_for_edge} and \ref{appendix_for_point} respectively. From these formulas we express the area of intersection in terms of $\theta$ using standard formulas for the area of polygons and circle sectors. This gives us formula $f_1(\theta)$ for the area of intersection for $0 \leq \theta \leq \arccos(\frac{2}{3}) - \frac{\pi}{6}$ and formula $f_2(\theta)$ for the area of intersection for $\arccos(\frac{2}{3}) - \frac{\pi}{6} \leq \theta \leq \frac{\pi}{6}$ (see Appendix~\ref{appendix_for_edge} and \ref{appendix_for_point}. By symmetry, we extend this to formula $f(\theta)$ for $0 \leq \theta \leq \frac{\pi}{3}$ given in Appendix~\ref{appendix_for_full_circle}.

\begin{figure}
\begin{center}
\includegraphics[width=2.2in]{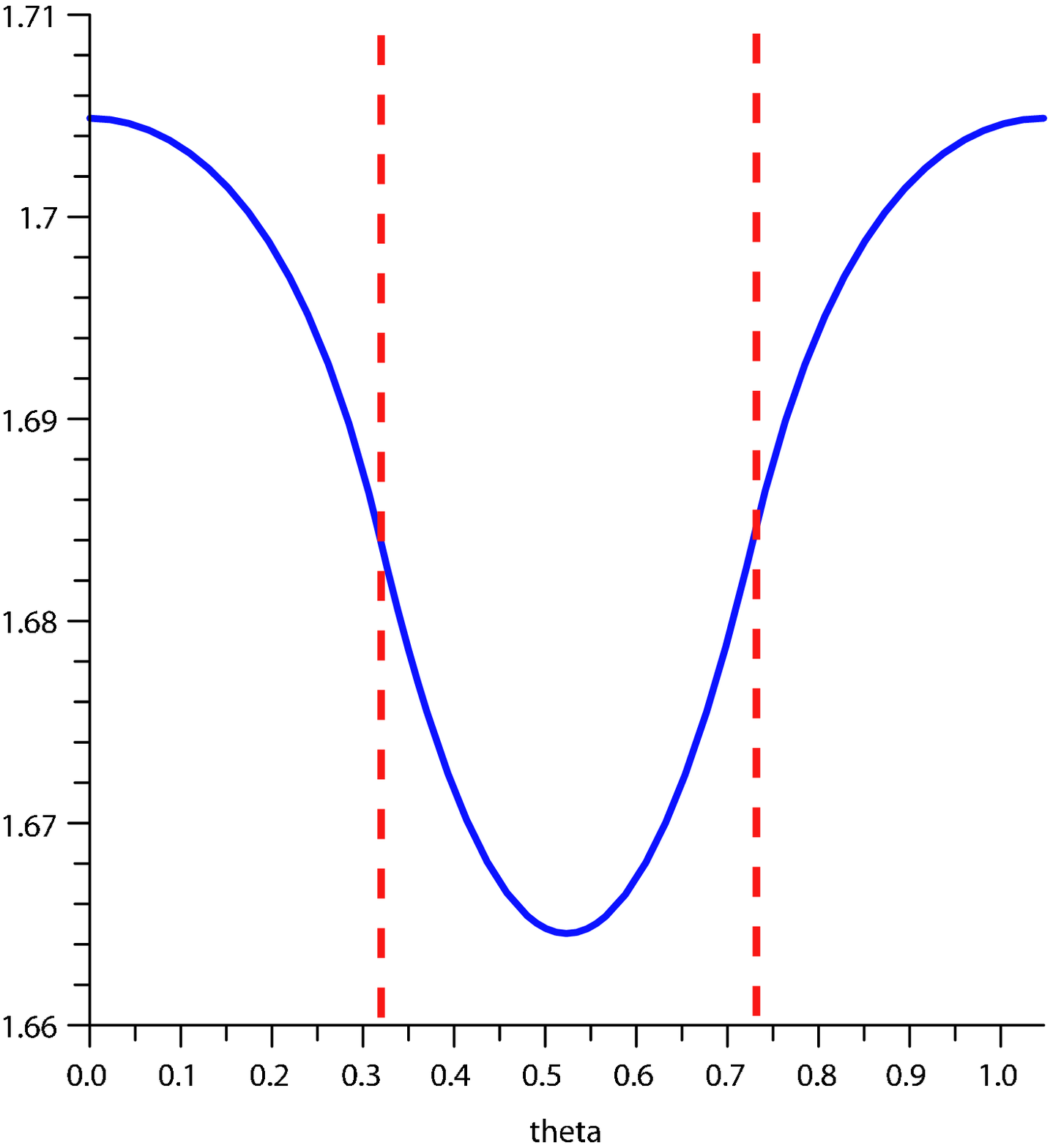}
\caption{Plot showing the area of intersection for $0 \leq \theta \leq \frac{\pi}{3}$.}
\label{intersectionareagraph}
\end{center}
\end{figure}

A plot showing $f(\theta)$ for the interval $0 \leq \theta \leq \frac{\pi}{3}$ is given in Figure \ref{intersectionareagraph}. Using Maple we find that $f'(\theta)$ (the first derivative of $f(\theta)$) is 0 for $\theta = \frac{\pi}{6}$. Thus, by symmetry, the minimum intersection occurs when $X$, the center point of $d$ and a vertex of of $h$ are collinear. Computing the value of $f(\theta)$ at any one of these points gives our value for $\Delta$, specifically

\begin{eqnarray*}
\Delta &=& {\frac{\sqrt{3}}{36}}\,+{\frac{\sqrt{11}}{12}}\,+{\frac{\pi}{2}}\, -{\frac{1}{2}}\,\arctan \left( {\frac{5\,\sqrt {3}-\sqrt {11}}{5+\sqrt {11}\sqrt {3}}} \right)\\
&\approx & 1.6645
\end{eqnarray*}

\onecolumn
\footnotesize
\subsection{Derived Formula for $0 \leq \theta \leq \arccos(\frac{2}{3}) - \frac{\pi}{6}$ (see Fig.~\ref{intersection_on_edge})}
\label{appendix_for_edge}

\begin{align*}
A_{x}(\theta) &= \frac{1}{3}\,\sqrt {3}-\frac{1}{2}\,\sqrt {1- \left( -\frac{2}{3}\,\sqrt {3}+\frac{1}{2}\,\sin \left(\theta \right) \sqrt {3}+\frac{1}{2}\,\cos \left( \theta \right)  \right) ^{2}}\sqrt {3}-\frac{1}{4}\,\sin \left( \theta \right) \sqrt {3}+\frac{3}{4}\,\cos \left(\theta \right)\\
A_{y}(\theta) &= 1+\frac{1}{2}\,\sqrt {1- \left( -\frac{2}{3}\,\sqrt {3}+\frac{1}{2}\,\sin \left( \theta\right) \sqrt {3}+\frac{1}{2}\,\cos \left( \theta \right)  \right) ^{2}}+\frac{1}{4}\,\sin \left( \theta \right) -\frac{1}{4}\,\cos \left( \theta \right) \sqrt {3}\\
B_{x}(\theta) &= \cos(\theta)\\
B_{y}(\theta) &= \sin(\theta)\\
C_{x}(\theta) &= \frac{1}{3}\,\sqrt {3}-\frac{1}{2}\,\sqrt {1- \left( \frac{2}{3}\,\sqrt {3}+\frac{1}{2}\,\sin \left(\theta \right) \sqrt {3}-\frac{1}{2}\,\cos \left( \theta \right)  \right) ^{2}}\sqrt {3}+\frac{1}{4}\,\sin \left( \theta \right) \sqrt {3}+\frac{3}{4}\,\cos \left(\theta \right)\\
C_{y}(\theta) &= -1-\frac{1}{2}\,\sqrt {1- \left( \frac{2}{3}\,\sqrt {3}+\frac{1}{2}\,\sin \left( \theta\right) \sqrt {3}-\frac{1}{2}\,\cos \left( \theta \right)  \right) ^{2}}+\frac{1}{4}\,\sin \left( \theta \right) +\frac{1}{4}\,\cos \left( \theta \right) \sqrt {3}\\
D_{x}(\theta) &= \frac{2}{3}\,\sqrt{3}\\
D_{y}(\theta) &= \frac{2}{3}\\
E_{x}(\theta) &= \frac{2}{3}\sqrt{3}\\
E_{y}(\theta) &= -\frac{2}{3}
\end{align*}

\begin{align*}
f_{1}(\theta) =&\,\frac{1}{2}\left( A_{x}(\theta)(B_{y}(\theta) - D_{y}(\theta)) + B_{x}(\theta)(C_{y}(\theta) - A_{y}(\theta)) + C_{x}(\theta)(E_{y}(\theta) - B_{y}(\theta)) + E_{x}(\theta)(D_{y}(\theta) - C_{y}(\theta)) + D_{x}(\theta)(A_{y}(\theta) - E_{y}(\theta)) \right)\\
& + \frac{1}{2}\left( \pi +\arctan \left( {\frac {- \left( -B_{x}(\theta)+C_{x}(\theta) \right)  \left(-A_{y}(\theta)+B_{y}(\theta) \right) + \left( B_{y}(\theta) -C_{y}(\theta) \right)  \left( A_{x}(\theta)-B_{x}(\theta) \right) }{ \left( -B_{x}(\theta)+ C_{x}(\theta) \right)  \left( A_{x}(\theta)-B_{x}(\theta) \right) + \left( B_{y}(\theta)-C_{y}(\theta) \right)  \left( -A_{y}(\theta)+B_{y}(\theta) \right) }} \right)\right)
\end{align*}

\subsection{Derived Formula for $\arccos(\frac{2}{3}) - \frac{\pi}{6} \leq \theta \leq \frac{\pi}{6}$ (see Fig.~\ref{intersection_on_point})}
\label{appendix_for_point}

\begin{align*}
A_{x}(\theta) &= \frac{1}{3}\,\sqrt {3}-\frac{1}{2}\,\sqrt {1- \left( -\frac{2}{3}\,\sqrt {3}+\frac{1}{2}\,\sin \left(\theta \right) \sqrt {3}+\frac{1}{2}\,\cos \left( \theta \right)  \right) ^{2}}\sqrt {3}-\frac{1}{4}\,\sin \left( \theta \right) \sqrt {3}+\frac{3}{4}\,\cos \left(\theta \right)\\
A_{y}(\theta) &= 1+\frac{1}{2}\,\sqrt {1- \left( -\frac{2}{3}\,\sqrt {3}+\frac{1}{2}\,\sin \left( \theta\right) \sqrt {3}+\frac{1}{2}\,\cos \left( \theta \right)  \right) ^{2}}+\frac{1}{4}\,\sin \left( \theta \right) -\frac{1}{4}\,\cos \left( \theta \right) \sqrt {3}\\
B_{x}(\theta) &= \cos(\theta)\\
B_{y}(\theta) &= \sin(\theta)\\
C_{x}(\theta) &= \frac{2}{3}\,\sqrt{3}\\
C_{y}(\theta) &= -\frac{1}{3}\,\sqrt {-3+12\,\cos \left( \theta \right) \sqrt {3}-9\, \left(\cos \left( \theta \right)  \right) ^{2}}+\sin \left( \theta \right)\\
D_{x}(\theta) &= \frac{2}{3}\sqrt{3}\\
D_{y}(\theta) &= \frac{2}{3}
\end{align*}

\begin{align*}
f_{2}(\theta) =&\,\frac{1}{2}\left( A_{x}(\theta)(B_{y}(\theta) - D_{y}(\theta)) + B_{x}(\theta)(C_{y}(\theta) - A_{y}(\theta)) + C_{x}(\theta)(D_{y}(\theta) - B_{y}(\theta)) + D_{x}(\theta)(A_{y}(\theta) - C_{y}(\theta)) \right)\\
& + \,\frac{1}{2}\left( \pi +\arctan \left( {\frac {- \left( -B_{x}(\theta)+C_{x}(\theta) \right)  \left(-A_{y}(\theta)+B_{y}(\theta) \right) + \left( B_{y}(\theta) -C_{y}(\theta) \right)  \left( A_{x}(\theta)-B_{x}(\theta) \right) }{ \left( -B_{x}(\theta)+ C_{x}(\theta) \right)  \left( A_{x}(\theta)-B_{x}(\theta) \right) + \left( B_{y}(\theta)-C_{y}(\theta) \right)  \left( -A_{y}(\theta)+B_{y}(\theta) \right) }} \right)\right)
\end{align*}

\subsection{Derived Formula for  $0 \leq \theta \leq \frac{\pi}{3}$}
\label{appendix_for_full_circle}

\begin{equation*}
f(\theta) = \begin{cases}
f_{1}(\theta)&\text{if } 0 \leq \theta < \arccos(\frac{2}{3}) - \frac{\pi}{6} \\
f_{2}(\theta)&\text{if } \arccos(\frac{2}{3}) - \frac{\pi}{6} \leq \theta < \frac{\pi}{6} \\
f_{2}(\frac{\pi}{3} - \theta)&\text{if } \frac{\pi}{6} \leq \theta < \arccos(\frac{2}{3}) \\
f_{1}(\frac{\pi}{3} - \theta)&\text{if } \arccos(\frac{2}{3}) \leq \theta \leq \frac{\pi}{3} \\
\end{cases}
\end{equation*}

\end{document}